# Electric Field Effect in Multilayer Cr$_2$Ge$_2$Te$_6$: a Ferromagnetic Two-Dimensional Material

Wenyu Xing [1,2,†], Yangyang Chen[1,2,†], Patrick M. Odenthal[3], Xiao Zhang[1,2], Wei Yuan[1,2], Tang Su[1,2], Qi Song[1,2], Tianyu Wang[1,2], Jiangnan Zhong[1,2], Shuang Jia[1,2], X. C. Xie[1,2], Yan Li[3], and Wei Han[1,2*]

[1]International Center for Quantum Materials, School of Physics, Peking University, Beijing 100871, P. R. China

[2]Collaborative Innovation Center of Quantum Matter, Beijing 100871, P. R. China

[3]Department of Physics and Astronomy, University of Utah, UT 84112, USA

[†]These authors contributed equally to the work

*Correspondence to: weihan@pku.edu.cn (W.H.)

The emergence of two-dimensional (2D) materials has attracted a great deal of attention due to their fascinating physical properties and potential applications for future nano-electronic devices. Since the first isolation of graphene, a Dirac material, a large family of new functional 2D materials have been discovered and characterized, including insulating 2D boron nitride, semiconducting 2D transition metal dichalcogenides and black phosphorus, and superconducting 2D bismuth strontium calcium copper oxide, molybdenum disulphide and niobium selenide, etc. Here, we report the identification of ferromagnetic thin flakes of Cr$_2$Ge$_2$Te$_6$ (CGT) with thickness down to a few nanometers, **which provides a very important piece to the van der Waals structures consisting of various 2D materials. We further demonstrate the giant modulation of the channel resistance of 2D CGT devices via electric field effect. Our results illustrate the gate voltage tunability of 2D CGT and the potential of CGT, a ferromagnetic 2D material, as a new functional quantum material for applications in future nanoelectronics and spintronics.**

**Introduction**

The research on graphene and post-graphene two-dimensional (2D) materials has been one of the most exciting research fields in the last decade [1-17]. Graphene is a Dirac material that has exhibited a lot of interesting/superior physical properties, including extremely high mobility, room temperature quantum Hall effect, and long spin lifetimes [1-4,13,17]. 2D boron nitride is an insulating material, which could be used a tunneling barrier and protecting layer for other 2D materials [7,8,10]. Semiconducting 2D transition metal dichalcogenides and black phosphorus could be very useful for nano electronics devices [6,7,9,14,16], such as field effect transistor with large on/off ratio. Superconducting 2D materials also provide interesting physical properties, such as Ising superconductivity [11,12,18]. Furthermore, the structures consisting of various 2D materials, van der Waals structures, have shown unusual properties and new phenomena [8,10]. Due to their ultra-thin nature, an electric field could give rise to the large modulation of the Fermi level, thus, providing a unique way to greatly modulate their physical properties. This electric field gating effect makes 2D materials very promising candidates for future nanoelectronics applications [9,19]. Recently, 2D ferromagnetic/antiferromagntic materials has also attracted a lot of interest [20-26]. Very intriguingly, Gong et al have demonstrated the 2D ferromagnetism in $Cr_2Ge_2Te_6$ (CGT) down to 2 layers using optical Kerr rotation [26]. However, for the 2D CGT, the electrical transport properties and the electric field effect have been lacking.

In this study, we report the preparation and characterization of ferromagnetic thin flakes of Cr$_2$Ge$_2$Te$_6$ (CGT) with thickness down to a few nanometers. Furthermore, large modulation of the channel resistance based on the 2D CGT flakes is demonstrated via the electric field effect. These results make CGT a very important ferromagnetic 2D materials candidate to the van der Waals structures consisting of various 2D materials [8,10,27].

**Experimental methods**

The CGT single crystal is a layered material belonging to the $R\bar{3}$ group. High-quality CGT single crystals are synthesized using the flux method from high-purity elemental Cr (99.99%), Ge (99.999%) and Te (> 99.999%) materials ordered from Alfa Aesar Company. A mixture of these materials with an atomic ratio of Cr: Ge: Te equal to 10: 13.5: 76.5, is sealed in evacuated quartz ampoules and heated at 1000 °C for 1 hour, followed by slow cooling to 450 °C for a period of 90 hours. The residual Ge and Te fluxes are removed by a centrifugation step at the end. The detailed chemical and structural properties of synthesized Cr$_2$Ge$_2$Te$_6$ single crystal can be found in our previous report [28].

The 2D CGT flakes are prepared using the well-established mechanical exfoliation method [29]. The SiO$_2$/Si wafers are first cut to be ~ 10 × 10 mm$^2$, and followed by an ultrasonical cleaning in acetone, IPA and DI water in sequence. Then the tape with CGT flakes is put directly onto the substrates and annealed for ~ 2 min at ~ 100 °C on a hot plate. This heating step has been shown to be helpful to synthesize large area graphene previously [30]. The tape is removed after the sample is cooled down and the 2D CGT flakes are transferred onto the SiO$_2$/Si substrates, which are identified by a Nikon high-resolution optical microscope.

For the magneto-optic Kerr effect (MOKE) measurement, optical microscopy is used to identify the location of the laser (spot size: ~ 15 μm$^2$) on the 2D CGT flakes. During the measurement, the linear polarized HeNe laser (wave length: 633 nm; intensity: 15 W/cm$^2$) is modulated in intensity at 84 kHz with a photoelastic modulator, and the incidence angle of the laser is 18 degrees from the normal direction. The beam intensity reflected off the CGT surface is measured with a Lock-in amplifier for better the signal-to-noise ratio.

The devices are fabricated on 2D CGT flakes using electron-beam lithography and the electrical contacts are made of ~ 80 nm Pt grown by magnetron sputtering. After a lift-off process with acetone to remove the electron-beam resist consisting of PMMA/MMA bilayers, Al wires are bonded to the large Pt contact pads for electrical measurements. The devices are measured in an Oxford Spectromag system with temperature from 300 to 1.5 K using the Keithley K2400 source meters and K2002 voltage meters.

**Results and discussion**

Preparation of the few-layer Cr$_2$Ge$_2$Te$_6$

As illustrated in Fig. 1a, along the c-axis, a layer of Cr atoms is sandwiched between two Ge-Te layers, which forms a single layer of CGT. These layers are stacked together via the van der Waals interaction with an interlayer distance of 6.9 Å, thus, single layer and few-layer CGT could be prepared by the mechanical exfoliation method [8,29,31]. The ferromagnetic properties of the synthesized CGT single crystal are characterized using a magnetic properties measurement system (Quantum Design MPMS-3). As temperature decreases below ~ 61 K in a magnetic field of 0.1 T along the crystal's *ab*-plane, an abrupt enhancement of the magnetization is observed (Fig. 1b), which indicates the Curie temperature of ~ 61 K. At 5K, the magnetization of CGT single crystal reaches saturation at the value ~2.4 $\mu_B$ per Cr atom (inset of Fig. 1b), which is close to the expected value for the high spin configuration state of $Cr^{3+}$ (3 $\mu_B$ per Cr atom). As the temperature decreases, the perpendicular magnetic anisotropy increases. The detailed characterization of the magnetic properties of the CGT bulk single crystal can be found in our earlier report [28].

Using the heating-assisted mechanical exfoliation method [30], we are able to achieve the 2D CGT flakes of various thicknesses down to a few nanometers. Fig. 1c-1e show the optical images of three typical CGT flakes, of which a brighter color corresponds to a thicker film. The thicknesses of these flakes are determined by atomic force microscopy (AFM) using the tapping mode (see supplementary Fig. S1a-S1c). As shown in Fig. 1f-1h, the thicknesses for the CGT flakes are determined to be ~ 4.5 nm, ~ 8.5 nm, and ~ 40 nm respectively, through height averaging in the line cuts, which are indicted by white dashed lines in Fig. 1c-1e.

Ferromagnetic properties of thin $Cr_2Ge_2Te_6$

The magneto-optic Kerr effect (MOKE) is utilized to characterize the ferromagnetic properties of the exfoliated 2D CGT flakes. Fig. 2a-2d show the magnetic field dependence of the Kerr rotation in arbitrary units for a ~ 40 nm CGT flake at 3, 20, 40, and 60 K, respectively, after the subtraction of a linear background and normalization. During the measurement, the magnetic field is applied parallel and in the plane of the 2D CGT flake using the longitudinal orientation. The inset of Fig. 2a shows the optical image of this ~ 40 nm flake and the red dot indicates the location of the laser spot. At 3 K, a pronounced magnetic hysteresis loop is observed, which unambiguously demonstrates the ferromagnetic property of the ~ 40 nm CGT flake. The mild yet distinguishable hysteresis loop of the Kerr rotation at 60 K indicates the Curie temperature is higher than 60 K, which is close to the bulk Curie temperature of ~ 61 K [28,32]. This can be explained by the relative thick nature of this CGT flake. We are not able to obtain the MOKE hysteresis on ~ 20 nm CGT flake (see supplementary information S1 and Fig. S2). One possible reason is related to the limited sensitivity of our current MOKE setup (~1 micro radian), and the other possible cause is the oxidation of the surface layer of the 2D CGT. Since the monolayer CGT is predicted to be ferromagnetic with a Curie temperature of 106 K [33], the MOKE measurement on the thin CGT samples down to single layer requires future studies using high-resolution MOKE, such as Sagnac interferometer enhanced magneto-optic measurements that can reach several nano radian sensitivity and was previously used to probe the broken time reversal symmetry in superconducting $Sr_2RuO_4$ [34].

Electric field effect of the thin $Cr_2Ge_2Te_6$ flakes

One major advantage of the 2D materials is that their physical properties could be largely tuned via the modulation of the Fermi level by a perpendicular electric field [9,31,35]. To explore the electric field effect in the 2D CGT, we fabricate Hall-bar devices using the nanofabrication technique. To avoid degradation due to oxidation (see supplementary S2 and Fig. S3) [21], we tried to minimize the time that CGT is exposed in air during fabrication and the devices are measured in vacuum. Fig. 3a shows the schematic drawing of the device fabricated on the 2D CGT flakes on the $SiO_2$/Si substrate, where the thickness of $SiO_2$ is ~ 300 nm and the highly electron-doped Si is used as a back gate to provide the electric fields. Three representative devices fabricated on different thicknesses of CGT flakes are selected to describe the main results in our study. Device A, B, and C are fabricated on ~ 25 nm, ~ 15 nm, and ~ 12 nm thick CGT flakes, respectively, where the thicknesses are determined by AFM measurement after the electrical measurements. The channel resistances are obtained by the four-probe measurement in an Oxford Spectromag system. Fig. 3b shows the typical voltage ($V_{xx}$) vs. the current ($I$) measured at 300 K for Device A made on ~ 25 nm CGT (channel width/length: 1.7/2.5 μm) at the gate voltages of 80, 40, 0, -40, -80 V, respectively. The gate voltage dependence of the channel resistances that are extracted from the slope of the $V_{xx}$ vs. $I$ curves is summarized in Fig. 3c. A more dramatic modulation of the channel resistances is observed on Device B with ~ 15 nm CGT (channel width/length: 1.2/1.5 μm; see supplementary Fig. S4), and Device C with ~ 12 nm CGT (channel width/length: 1.7/0.8 μm), as shown in Figs. 3d and 3e. Fig. 3d shows the typical $V_{xx}$ vs. $I$ curves measured at 300 K for Device C at the gate voltages of 0, -20, -40, -60, and -80 V, respectively, and the channel resistance vs. gate voltage is plotted in Fig. 3e. The maximum resistance of 117 MΩ at $V_g$ = 80 V is ~ 50 times larger than the lowest value of 2.2 MΩ at $V_g$ = -100 V. This gating effect is much more profound in Device C compared to Device A. The universal feature of less gating effect in thicker CGT films is further confirmed by measurements on Device D (~ 29 nm) and E (~ 18 nm) (see supplementary information S3 and Fig. S5). For even thinner flakes less than ~ 8 nm, we have been unable to contact the samples for electrical measurements even under large gate voltages (see supplementary information S4 and Fig. S6), which we believe might be due to similar issues, including large contact resistances and oxidation during the fabrication process (See supplementary Fig. S3), as observed on semiconducting 2D materials, such as transition metal dichalcogenides [36-38].

The channel resistances of these devices are measured as a function of temperature under different gate voltages (Fig. 4a-4c). For Device A with ~ 25 nm CGT (Fig. 4a), the ungated device exhibits an semiconducting feature, which is consistent with our previous study on bulk CGT [28]. The gate tuning of the channel resistance shows the same trend as that at 300 K in the full temperature range, but the modulation is more dramatic at low temperatures. When the films become thinner, the channel resistances exhibit a large increase. For instance, at 300 K and $V_g$ = 0 V (blue squares in Fig. 4b-c), Device B (~ 15 nm CGT) has a channel resistance of ~ 10 MΩ, and Device C (~ 12 nm CGT) shows channel resistance of ~ 130 MΩ. For Device B, the gate voltage modulation of the channel resistance is more profound compared to Device A. At 1.5 K, the ratio of maximum/minimum in channel resistance is ~ 70 for Device B, which is much larger compared to Device A. For even thinner flakes, a giant modulation of the channel resistance is observed. As shown in Fig. 4c for Device C, the channel resistance changes from very insulating (not measurable below 250 K) at 0 V to quite conducting, which shows a resistance of ~ 1.6 MΩ at -100 V and 1.5 K. At low temperatures, a non-linear feature of the $V_{xx}$ vs. $I$ curves (Fig. S6) for thinner CGT devices could be associated to the existence of the hopping transport mechanism. The resistance values are obtained from the linearly fitted lines at the high bias region (see supplementary information S5 and Fig. S7). From the temperature dependence of the channel resistance, two major conclusions could be drawn. First, for the electrical properties of the CGT flakes, a clear thickness dependence is observed. As the thickness decreases, the resistivity increases, and the gate effect is more profound. This observation can be explained by the distribution of the induced charge carriers over the layers when the CGT is thicker. Second and most importantly, for thicker CGT Device A under -80 V, the channel resistance increases as the temperature decreases, still indicating a semiconducting feature. Meanwhile, for thinner CGT Device C, the channel resistance decreases as the temperature decreases under the gate voltage of

-80 V and -100 V (Fig. 4c inset), indicting a metal-like behavior, in stark contrast to the semiconducting/insulating properties while ungated.

The magnetic properties of the gated devices are characterized by anomalous Hall effect in a magnetic field applied perpendicular to the device geometry. Fig. 5 shows the measurement of the anomalous Hall resistance on Device B at -80 V and 5 K. A clear switching behavior is observed at ~ 0.2 T when the magnetic field is swept from -1 to 1 T (black line), and a sharp drop is observed at ~ -0.2 T while sweeping the magnetic field from 1 to -1 T (red line). These two sharp switching behaviors indicate the ferromagnetic properties of the 2D CGT when it is gated to be close to the ferromagnetic metallic region. The characterization of the ferromagnetic properties of the ungated devices based on anomalous Hall effect measurement is quite challenging due to the extremely resistive properties of CGT. Using bulk CGT for example, even the ordinary Hall measurement is hard to get a clear signal after the CGT becomes quite insulating at low temperatures (see supplementary Fig. S8). The fully understanding of the electric field effect on the magnetic anisotropy and Curie temperature requires future studies on the same devices by both MOKE and AHE.

**Conclusion**

In summary, we have succeeded in the preparation of few-layer 2D CGT, which exhibits ferromagnetic properties below the Curie temperature, and in the achievement of gate voltage tuning of the channel resistances for the thin CGT devices. Our results could lead to the future studies on the electric field tuning of the ferromagnetic properties, such as Curie temperature, magnetic anisotropy, of the 2D ferromagnetic CGT of various thickness, and on the van der Walls structures [8,10,27], such as CGT and graphene/topological insulator based Dirac materials for quantum anomalous Hall effect [32,39]. Furthermore, the optimization of the isolation technique, such as sandwiched between two BN layers [8,40-42], could give rise to a high quality single layer CGT, which is theoretically predicted to have a Curie temperature of 106 K [33].

**Note added:** during the review process of our manuscript, we noticed the identification of intrinsic long-range ferromagnetic order in pristine $Cr_2Ge_2Te_6$ atomic layers down to 2 layers via Sagnac interferometer enhanced magneto-optic measurements by Gong, et al [26].

**Acknowledgements**

We acknowledge the fruitful discussions with Professors Ji Feng, and Xi Lin. We acknowledge the financial support from National Basic Research Programs of China (973 program Grant Nos. 2015CB921104, 2014CB920902, 2013CB921901 and 2014CB239302) and National Natural Science Foundation of China (NSFC Grant No. 11574006). W.H. also acknowledges the support by the 1000 Talents Program for Young Scientists of China.

**Figure Captions**

**Figure 1 | Preparation of the ferromagnetic 2D $Cr_2Ge_2Te_6$ (CGT) from a single crystal. a**, Schematic drawing of the CGT's layered crystal structure along the (0001) orientation. These layers are stacked together via the van der Waals interaction with an interlayer distance of ~6.9 Å. One monolayer CGT is ~ 6.9 Å thick, indicated by the black arrow. **b**, The magnetization of CGT single crystal as a function of the temperature under a magnetic field of 0.1 T. The inset is the magnetic field dependence of the magnetization measured at 5 K. During the measurements, the magnetic field is applied parallel to the crystal's ab-plane. **c-e**, The optical images of three typical 2D CGT flakes with different thicknesses. **f-h**, The thicknesses of these three 2D CGT flakes determined from the AFM measurement (see supplementary Fig. S1a-S1c). These line cuts are measured along white dashed lines in the optical images shown in Fig. 1c-1e.

**Figure 2 | Magneto-optical Kerr effect measurement on a 2D CGT flake (thickness: ~ 40 nm). a-d**, The magnetic field dependence of the Kerr rotation on 2D CGT flake measured at 3, 20, 40, and 60 K, respectively, after the subtraction of a linear background and normalization. The red curves indicate the magnetic field swept from positive to negative, and the black curves indicate the magnetic field swept from negative to positive. The inset in Fig. 2a is the optical image of this 2D CGT flake, where the red dot indicates the location of the linear polarized HeNe laser.

**Figure 3 | Gate voltage tuning of the channel resistances of the 2D CGT devices at 300 K. a,** Schematic drawing of the Hall bar devices fabricated on the 2D CGT flakes on the SiO2/Si substrate, where the n-doped Si is used as a back gate, the thickness of the SiO2 layer is 300 nm

and Pt electrodes are used for electrical contacts. **b**, The four-probe measurement of the channel resistances of Device A (CGT thickness: ~ 25 nm; channel width/length: 1.7/2.5 μm) at various gate voltages at 300 K. **c**, The gate voltage dependence of the channel resistance for Device A at 300 K. **d**, The four-probe measurement of the resistance of Device C (CGT thickness: ~ 12 nm; channel width/length: 1.7/0.8 μm) at various back gate voltages at 300 K. **e**, The back gate voltage dependence of the channel resistances of Device C at 300 K.

**Figure 4 | Large modulation of the channel resistances of the 2D CGT devices by electric field . a-c,** The temperature dependence of the channel resistances at various gate voltages for Device A (CGT thickness: ~ 25 nm; channel width/length: 1.7/2.5 μm), B (CGT thickness: ~ 15 nm; channel width/length: 1.2/1.5 μm), and C (CGT thickness: ~ 12 nm; channel width/length: 1.7/0.8 μm), respectively. Inset of Fig. 4c: The temperature dependence of the channel resistances at -80 and -100 V, which exhibits metal-like behavior.

**Figure 5 | Anomalous Hall measurement on the gated 2D CGT device (Device B).** The magnetic field dependence of the Hall resistance ($R_{xy}$) for gated Device B at -80 V and 5 K after the subtraction of a linear background. The red curve indicates the magnetic field swept from positive to negative, and the black curve indicates the magnetic field swept from negative to positive. The two large resistance changes at ~ 0.2 T and ~ -0.2 T are the signatures of the ferromagnetic properties of the gated Device B, and ~ 0.2 T corresponds to the coercive field of the gated 2D CGT.

Figure 1

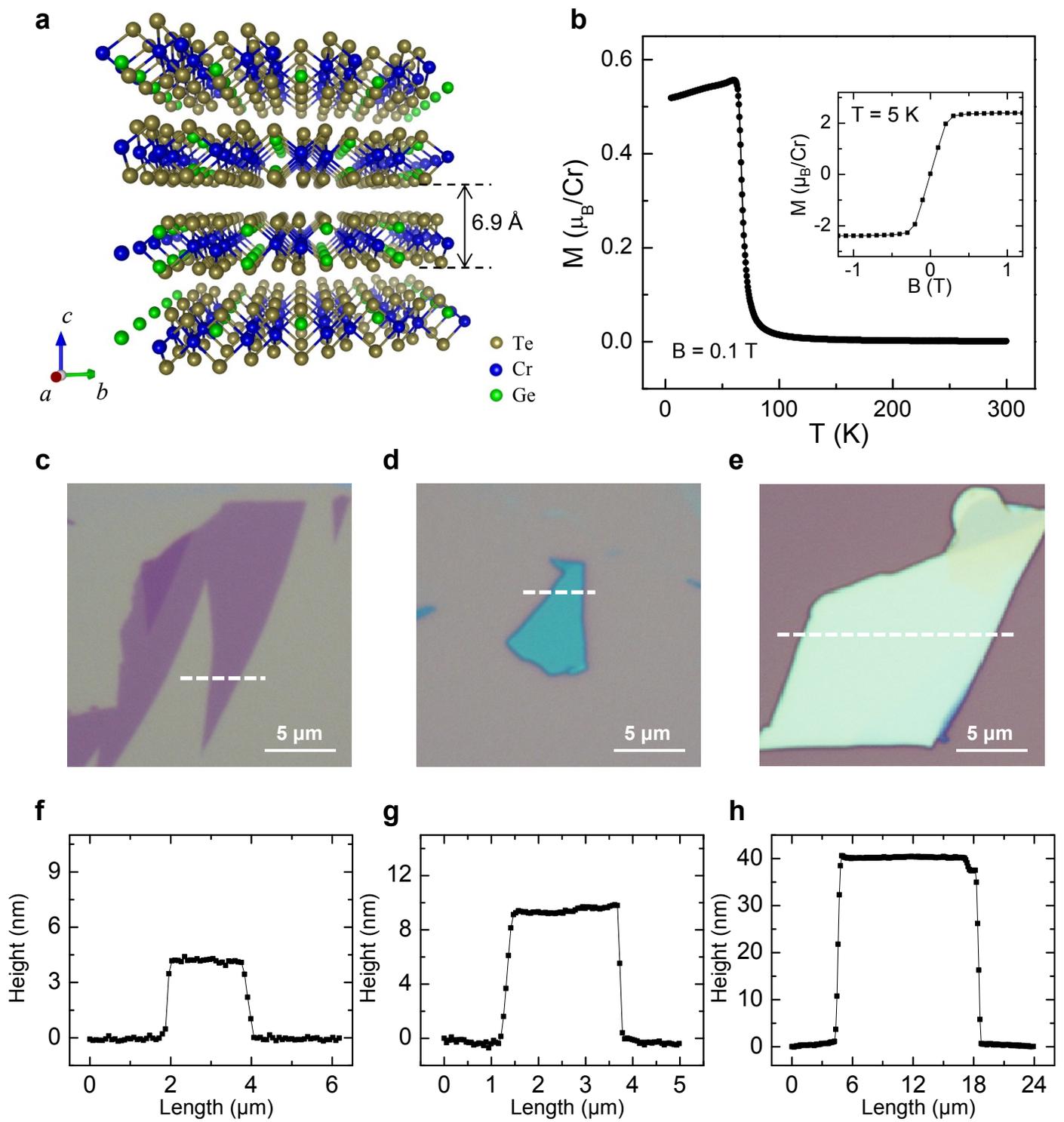

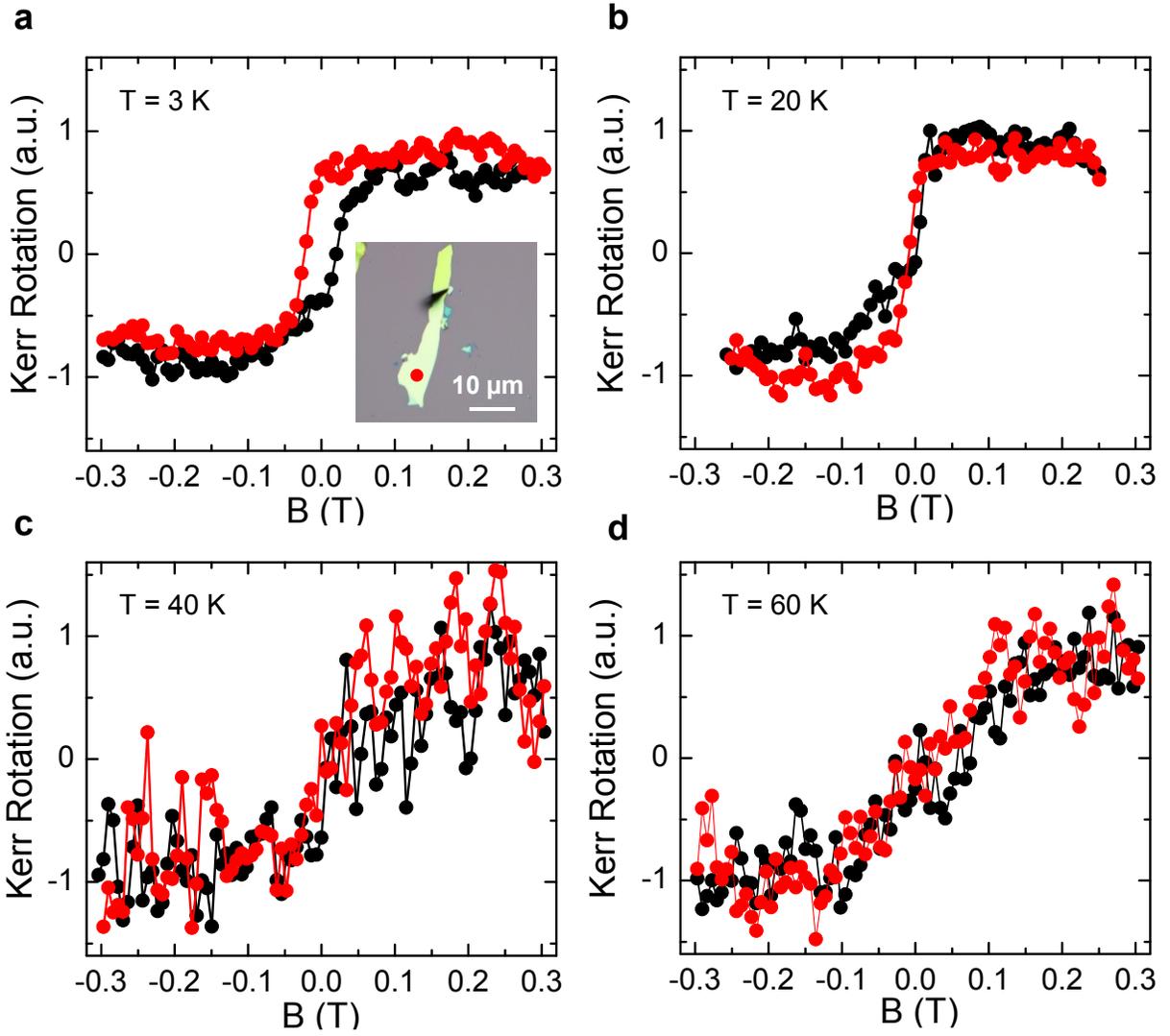

Figure 3

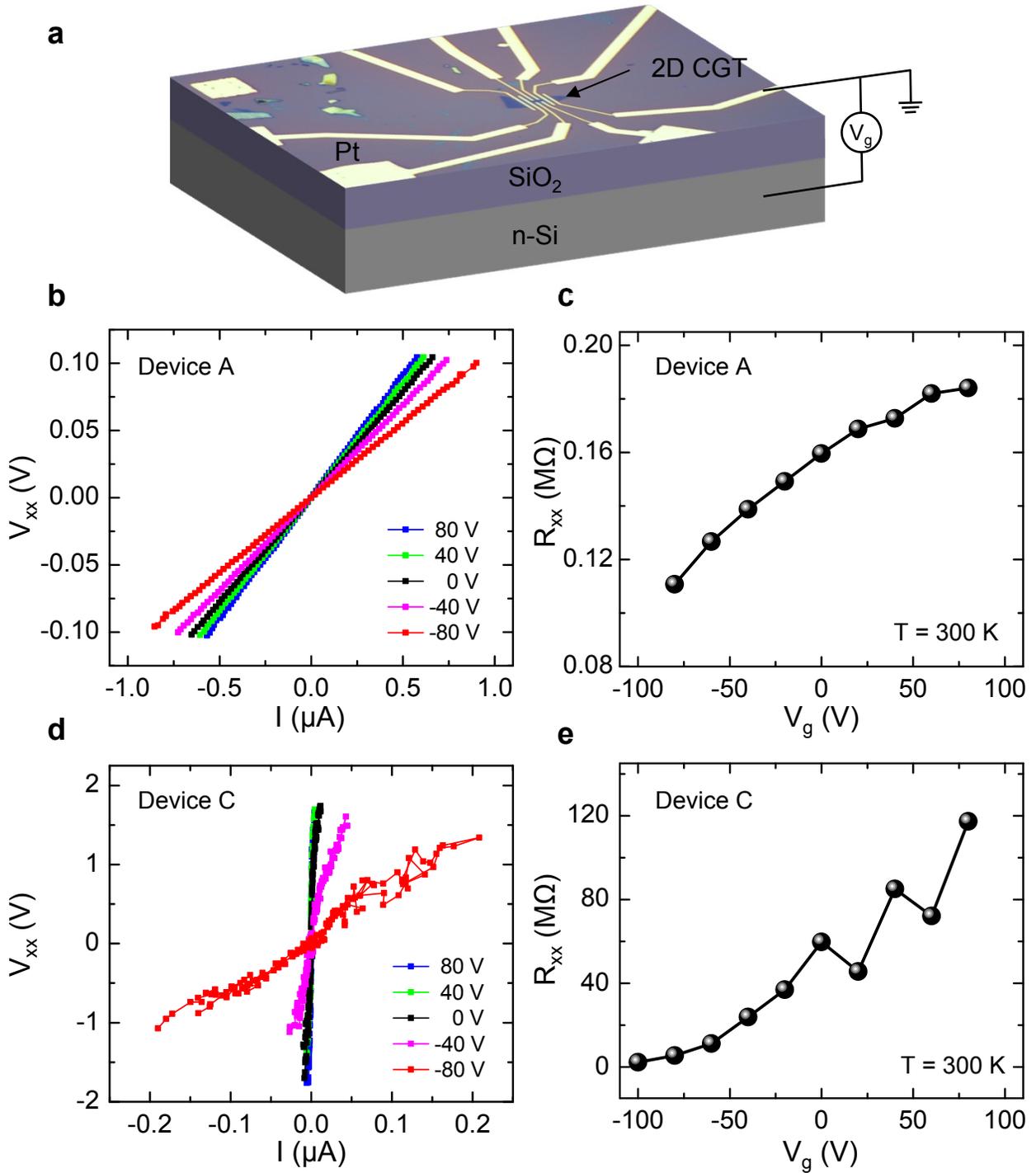

Figure 4

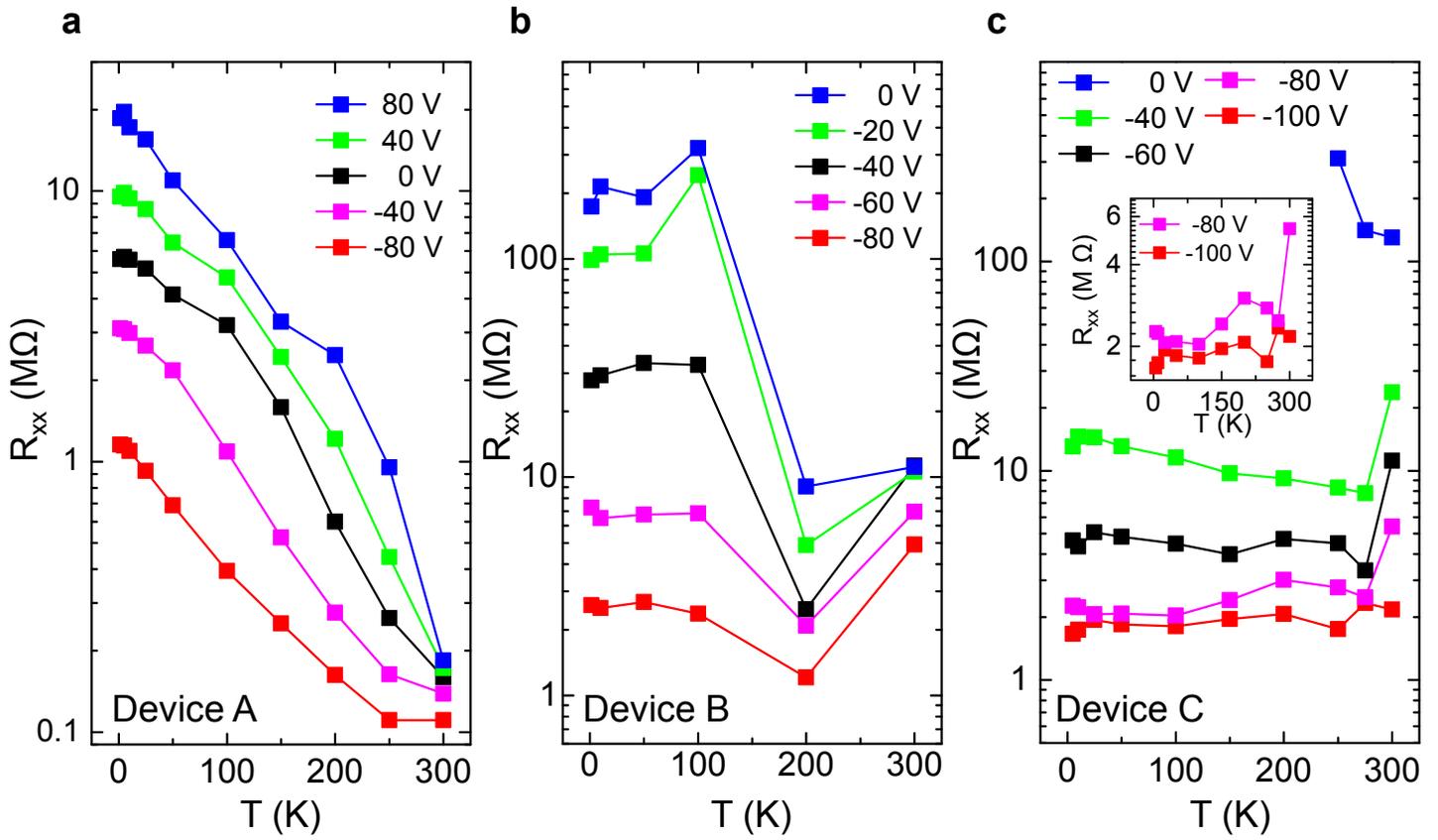

Figure 5

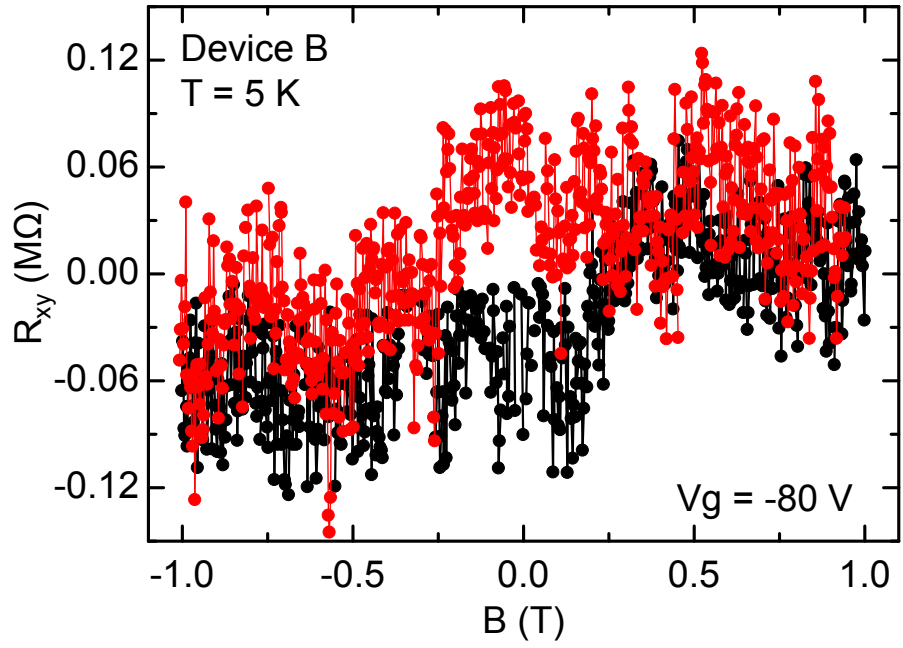